\documentclass[a4paper,12pt]{article}

\def\ad   {a^{\dagger}}

\def\HP   {\hat {P}}

\def\al   {\alpha}

\def\HH {\hat H}

\def\HJ {\hat J}

\def\HP {\hat P}

\def\HH {\hat H}

\def\oc {\overline c}

{\it}
{\it}
{\it}
{\it}
{\it}
{\it}

\def\de {\delta}

\def\eps {\epsilon}

\def\om {\omega}
\def\Om {\Omega}

\usepackage{times}
\usepackage{graphicx}
\begin{document}
\title{ Description of nuclei around $N=20$ starting from the Argonne V18 interaction.}
\author{G. Puddu\\
       Dipartimento di Fisica dell'Universita' di Milano,\\
       Via Celoria 16, I-20133 Milano, Italy}
\maketitle
\begin {abstract}
 Using the Argonne V18 interaction, renormalized with the Lee-Suzuki method,
 we study nuclei around the $N=20$ island of inversion. We include 5 major
 oscillator shells, in a no-core approach, using the Hybrid Multi-Determinant method reaching up to 
 few hundreds Slater determinants. Although qualitatively in agreement with the experimental
 levels, the calculated BE2 do not show the same amount of collectivity seen
 experimentally. 
\par\noindent
{\bf{Pacs numbers}}: 21.60.De, 21.10.-k$\,\,$  27.30.+t
\vfill
\eject
\end{abstract}
\section{ Introduction.}
 The evolution of the shell structure and magic numbers for extreme $N/Z$ ratios 
 has become one of the major topic both in experimental and theoretical nuclear physics (for a review
 see for example ref.[1]). Magic numbers may not be the same as we move away from  the stability line.
 The first instance was found in ref. [2] where it has been shown that the experimental data for $Na$ isotopes
 was inconsistent
 with the $N=20$ shell closure. In ref. [3] it has been  predicted that $Na$ isotopes  for $N\approx 20$ develop 
 large deformation  and that the orbitals coming from the $f$ shell becomes occupied (the basis used in ref.[3]
 is a deformed harmonic oscillator basis).
 Since then, many experimental and theoretical studies in this island of inversion region have been
  performed (cf. ref.[1]). 
 Among the others, shell model calculations have been performed using the (sd) and (fp) spherical major shells , ref.[4],
 and Monte Carlo Shell Model (MCSM) of ref.[5]. In these calculations an inert core has been assumed and a realistic
 effective interaction 
 has been adapted to the region (not to be confused with the renormalized interactions discussed below).

 To the author knowledge, no calculations starting from a  realistic bare nucleon-nucleon interaction have been performed
 for the island of inversion around $N\approx 20$.
 In this work we consider the Argonne V18 interaction (ref. [6]) and study $Ne$, $Mg$ and $Si$
 isotopes with neutron number
 $N=18, 20, 22$. Experimentally, Neon and Magnesium display the disappearance of the neutron shell
 closure at $N=20$, however Silicon
 at $N=20$ and heavier isotones have an increased excitation energy of the $2^+_1$ state,
 compared to the neighboring isotopes, pointing to a  restoration of the neutron shell closure.  
 Our approach is based on the Lee-Suzuki (LS) (refs.[7]-[10]) renormalization prescription, whereby the bare NN interaction
 is replaced by an effective interaction adapted to the large no core shell model space (cf. also ref.[11]).
 We consider only
 the effective two-body interaction and ignore the  many-body interactions induced by the renormalization 
 prescription. We limit our study to $5$ spherical major harmonic oscillator shells, although the real features of
 interaction are seen only with a large number of major shells. Once we determine the effective
 interaction, we use the Hybrid-Multi-Determinant method (HMD) (ref. [12]-[15]) to expand the 
 nuclear wave functions in terms
 of the most generic Slater determinants (i.e. no special symmetries are imposed). The results we obtained
 are mixed. From one hand the trend of the experimental excitation energies of the first $2^+$ (ref.[16])
 is reproduced, that
 is the energy of the $2^+_1$  decreases from $N=18$ to $N=22$ for $Ne$ and $Mg$ isotopes and it increases
 at $N=20$ for $Si$, however a good quantitative agreement is lacking. Also the BE2 are too small compared
 to the experimental data (cf. ref.[17] for a recent compilation), pointing to an insufficient
  collectivity for $Ne$ and $Mg$. 
As discussed in the 
 next sections, we evaluated also the average occupation numbers of the neutrons in the various orbits.
 The occupation numbers we obtained are very different from the ones obtained with realistic effective interactions.
 This is probably due to the very different nature of the effective interactions used in our context and
 in the MCSM and shell model context. This can be understood by considering that an increase of the single-particle
 space in the LS renormalization scheme, would make the LS effective interaction harder and harder at short
 distances. A "hard" interaction would scatter nucleons to all majors shell of the single-particle
 space, while in shell model calculations at the most one considers the $sd$ and $fp$ major shells.
\par
  Also we do not renormalize the transition operators nor
 any other. Following the results of ref. [18], we renormalize only operators that are very strong
 at short distances. We do not take the translational invariant quadrupole operator, rather we take the
 usual one in the lab frame with bare charges. The use of the laboratory frame operator
 does not lead
 to spurious center-of-mass contributions as long as one works
 with wave functions
 which factorize the intrinsic and $0\hbar\omega$ center-of-mass
 components (cf. the appendix of ref.[18]).
\par
 As an alternative method to the one used in this work we mention the ab-initio
 coupled-cluster method of ref.[19] and the in-medium similarity renormalization
 group method of ref. [20]. 
 We should mention that although we approximate the nuclear wave function with a
 linear combination of many deformed Slater determinants, we ultimately work with
 a spherical harmonic oscillator basis. Recently a different spherical basis has
 been proposed in ref. [21] as an alternative to the harmonic oscillator basis,
 i.e.  the Coulomb-Sturmian basis. In ref. [21] it has been argued that this basis  is better
 suited to the description of quantities (such as radii and quadrupole moments) which are
 sensitive to the tail of the single-particle wave functions.
\par
 The outline of this paper is the following. In section 2 we recall the HMD method. In section 3
 we discuss the results. In section 4 we give some concluding remarks.
 Since there are several ways that have been used to renormalize the NN interaction
 we give in the appendix a detailed description of the method we have used.
\bigskip
\section{ A brief recap of the method.}
\bigskip
       The HMD method (ref.[12]-[15]) is a variational approach to obtain eigenfunctions of an Hamiltonian.
       Given a spherical basis of $N_s$ single-particle states (e.g. an harmonic oscillator basis)
       the Hamiltonian is written as
$$
\HH = {1\over 2} \sum_{ijkl} H_{ij,kl}\ad_i\ad_j a_l a_k
\eqno(2.1)
$$
       where $ijkl$ label single-particle states ($i=1,2,...N_s$) and the one-body part has been included
       in the two-body interaction. We antisymmetrize from the start the matrix elements of the Hamiltonian
       $H_{ij,kl}=-H_{ij,lk}$, since exchange contributions are the opposite of direct terms. We describe eigenstates
       as a linear superposition of Slater determinants of the most generic type
$$
| \psi>= \sum_{S=1}^{N_D} g_S \HP |U_S>
\eqno(2.2)
$$
       where $\HP$ is a projector to good quantum numbers (e.g. good angular momentum and parity)
       $N_D$ is the number of Slater determinants $|U_S>$ expressed as
$$
|U_S> = \oc_1(S)\oc_2(S)... \oc_A(S) |0>
\eqno(2.3)
$$
       the generalized creation operators $\oc_{\alpha}(S)$ for $\al=1,2,..,A$ are a linear combination
       of the creation operators $\ad_i$
$$
\oc_{\al}(S)=\sum_{i=1}^{N_s}U_{i,\al}(S)\ad_i  \;\;\;\;\;\al=1,...A
\eqno(2.4)
$$
       The complex coefficients $U_{i,\al}(S)$ represent the single-particle wave-function of the
       particle $\al=1,2,..,A$. We do not impose any symmetry on the Slater determinants (axial or other)
       since the $U_{i,\al}$  are variational parameters. 
       These complex coefficients are obtained by minimizing the energy expectation values
$$
E[U]= { <\psi |\HH |\psi> \over <\psi |\psi>}
\eqno(2.5)
$$
       The coefficients $g_S$ are obtained by solving the generalized eigenvalue problem
$$
\sum_{S} <U_{S'} |\HP\HH | U_S> g_S = E \sum_{S} <U_{S'}|\HP| U_S> g_S  
\eqno(2.6)
$$
       for the lowest eigenvalue $E$.
\par
       A few comments are in order about this method. If we expand eq.(2.3) in terms of the spherical 
       single-particle creation operators, we include all possible contributions of the 
       spherical basis of the Hilbert space. However the coefficients of these contributions
       are built from products of the $U_{i,\al}$, and this is the reason why we must consider a large number of 
       Slater determinants $N_D$,
       even several hundreds. Second, the projector to good quantum numbers ideally should be
       the exact one (see for example ref.[22]) in terms of integrals over the Euler angles.
       However, our experience with the exact 3-dimensional projectors, tells us that the number of mesh points
       of the Euler angles has to be rather large, making the numerical integration computationally
       very expensive. If we were looking for the best approximation to eigenstates in terms of
       few Slater determinants, this would be unavoidable (actually in such a case we would prefer
       quasi-particle states), but we are looking for a sequence of states which
       approximate better and better the exact eigenstates. In order to make the calculations
       feasible, we prefer to consider projectors to good parity and z-component of the angular momentum.
       This way we can generate hundreds of Slater determinants. The resulting approximation for an eigenstate
       has the lowest energy for a given $J_z$ value. This is appropriate for low energy eigenstates
       of even-even systems. The method would be inadequate for odd-even  and odd-odd systems.
       In these cases we prefer to add to the Hamiltonian a term $\gamma \HJ^2$, with $\gamma>0$, 
       so that all states
       with $J\neq J_z$ would be moved to high energy and all states with $J< J_z$ are canceled by the $J_z$
       projector. This is particularly useful if we have
       a large single-particle basis and large number of particles. In order to improve the
       wave-functions, at the end of the calculation we always reevaluate the energies or transition
       probabilities replacing the $J_z$ projector with the exact three-dimensional projector to good angular
       momentum and parity. In the case of ${}^{33}Mg$ we use $\gamma\neq 0$. For large systems, the full angular
       momentum projector has to be nearly exact, otherwise the variational  method breaks down.
       In the final reprojection phase, instead, the number of mesh points for the integration over the Euler
       angles does not need to be very large. For the nuclei considered in this work a mesh
       of $24\times 12\times 24$ points is very accurate.
\par\noindent
       We consider a quasi-Newtonian minimization method. It is a generalization of
       the Broyden-Fletcher-Goldfarb-Shanno (BFGS) method (cf. for example ref.[23] and references in there). 
       The variant we use is described in detail in ref. [24].  
\par
       The essential idea of quasi-newtonian methods is the following. Consider for example the case $N_D=1$ that is,
       a single Slater determinant. The total number of variables is $2\times A\times N_s$, which are
       the real and imaginary parts of the $U_i,\al$. These can be assembled into a vector of components $x_n$
       and the energy is a function of this vector $E(x)$. We evaluate the energy gradient $g_n=\partial E\partial x_n$.
       We also need a search direction in the $x$-space, $s_n$. 

       At the iteration number $k$ let we update $x^k= x^{k-1}+a s^{k-1}$, where $a$ is a real number
       (the step of descent).
       We optimize $a$ so that $E(x^k)<E(x^{k-1})$ and evaluate $g^k$. We also evaluate an approximate
       inverse Jacobian matrix $G^k$ and update the search direction $s^k_n =-\sum_m G^k_{nm} g^k_m$. 
       Initially we take $G=1$ and hence $s=-g$. The approximate inverse Jacobian must satisfy
       the relation $G^k (g^k - g^{k-1} ) = x^k - x^{k-1}$, in a matrix notation. The differences among the various
       quasi-Newtonian methods consist in different ways of finding the matrix $G$. Note that this is equivalent of
       finding an array that satisfies a linear system, rather than finding a vector that satisfies a linear system
       (cf. ref.[24] for the details). 
\par
       Using this method, when we have several Slater determinants, we vary one Slater determinant at a time,
       not all Slater determinants simultaneously. The reason is that with this method we determine
       a direction and a step of descent in the energy hyper-surface. However  Slater determinants
       have different step sizes and therefore a simultaneous variation of all Slater determinants
       might need small step sizes, slowing down the convergence to the energy minimum. 
\par
       The minimization strategy consists in the following steps. We first generate the Hartree-Fock
       solution ($N_D=1$) then we add one Slater determinant at a time and optimize the last Slater
       determinant in order to minimize the energy. After we reach a specified number of Slater  
       determinants we vary all Slater determinants anew. We repeat this addition+refinement step
       several times. Each time increasing $N_D$ by several Slater determinants. The "addition" step
       can be non-trivial, we usually start with a very approximate Hartree-Fock, initially switch off the
       inverse Jacobian (this is the steepest descent method) and use huge values of the step sizes before
       using the quasi-newtonian method.
        For example, for $N_D=2$ we start
       from a very accurate Hartree-Fock (properly projected) and a very  approximate Hartree-Fock
       Slater determinants. We vary this very  approximate Hartree-Fock so as to minimize the energy
       and obtain a Slater determinant  $|U_2>$. Next we vary $|U_1>$ anew and then again
       $|U_2>$. We repeat several times until the energy does not decrease appreciably. We then add again 
       to the set of Slater determinants, an approximate Hartree-Fock (now $N_D=3$) and vary again the 
       last Slater determinant. We do not need to reoptimize all Slater determinant once we increase $N_D$
       by one. We usually do it once the set reaches certain numbers (typically $ 2,5,10,15,25,35,50,70,100,..$)     
\par
       The energies obtained with this method approach more and more, but are not, the exact energies.
       This however is not a severe problem. We are mostly interested in excitation energies.
       In the region of the island of inversion, the excitation energy of the first $2^+$ gives an indication
       of the breaking of shell closure. Both $E(0^+)$ and $E(2^+)$ approach the exact values from 
       above and  contain comparable errors of the same sign which mostly cancel in the difference.
       That is, excitation energies converge much faster than the energies themselves.
\par
       For many-particle systems the nucleon-nucleon interaction is very strong at small distances
       (or large relative momenta) and therefore the interaction must be renormalized. There are
       many prescriptions for the renormalization which necessarily give different effective
       Hamiltonians. We add in the appendix a detailed description of the adopted renormalization 
       procedure based on the Unitary-Model-Operator Approach (UMOA) of refs.[7]-[10], since
       this method has been implemented
       in several ways, and discuss a numerical test.
       In the next section we shall discuss the results obtained for ${}^{30,32,33,34}Mg$,
       ${}^{28,30,32}Ne$ and ${}^{32,34,36}Si$ isotopes using  $5$ major harmonic 
       oscillator shells.
\section{ Excitation energies around the $N=20$ island of inversion.}
\bigskip
       Our main focus is on the excitation energy of the first $2^+$ state for
       $Z=10,12,14$ around $N=18,20,22$. We consider only one harmonic oscillator frequency
       $\hbar\om=12$MeV. The NN interaction is the Argonne $v18$ potential (ref.[6]). 
      In all cases we add to the renormalized
       Hamiltonian  a center of mass term $\beta ( H_{cm} -3 \hbar \om /2)$ with 
       $\beta=1$, $H_{cm}$ being the center of mass harmonic oscillator Hamiltonian, in order
       to prevent spurious center of mass excitations.
       Typical size of the Hilbert space are $10^{29}\div 10^{30}$.
       For all nuclei we build in sequence wave functions consisting of 
      $N_D= 1,2,5,10,15,25,35,50,70,100,150,200,..$  Slater determinants (these numbers are somewhat 
      arbitrary). Every time we reach
       the above numbers  we re-optimize anew all Slater determinants.
       The energies require a very large number of Slater determinants to converge. Consider for
       example the energies of the first $0^+$ and of the first $2^+$ for ${}^{32}Mg$ shown in
       fig.1.
\renewcommand{\baselinestretch}{1}
\begin{figure}
\centering
\includegraphics[width=10.0cm,height=10.0cm,angle=0]{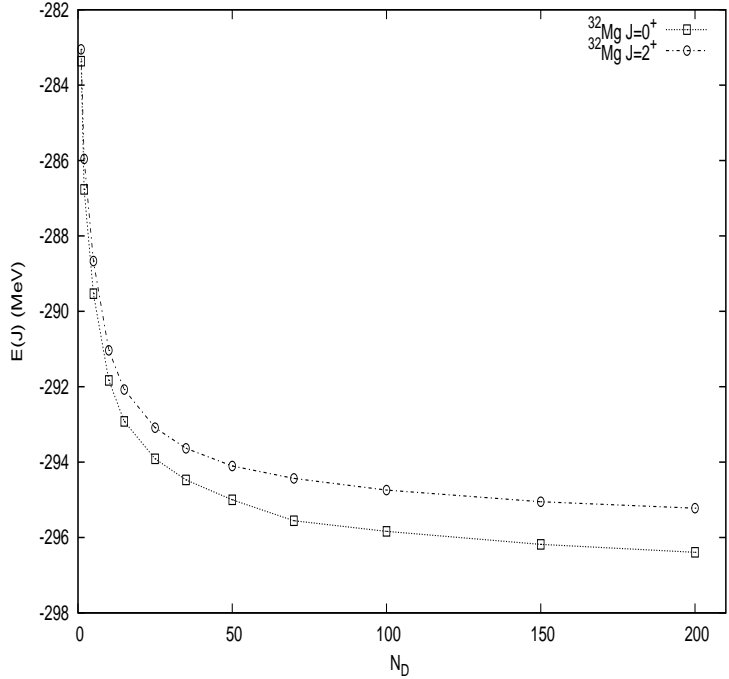}
\caption{Energies of ${}^{32}Mg$ for $J=0^+$ and $J=2^+$ as a function of the number of Slater determinants.}
\end{figure}
\renewcommand{\baselinestretch}{2}
  Although the pattern of convergence for both states is the same, the energies are themselves
       not fully converged. A very rough estimate (assuming an approximate $1/N_D$ behavior for the energies)
       indicates that the ground state is about $1\div 1.5$MeV lower.
       From fig.1 we can see that $5$ major shells do not describe well the energies: the binding energies
       are too large. Improved values can be obtained by increasing the number of major shells
       (the interaction is renormalized to a specific number of major shells). However
       calculations with $6$ or $7$ major shells are computationally much more involved 
       (for $6$ major shells we have $112$ single-particle states for both neutrons and protons and
       for $7$ major shells this number becomes $168$).
       In fig.2, we show the convergence of the excitation energy  for $Ne$ isotopes.
       The $N=20$ shell closure is absent, although not to the same extent seen experimentally.
       In fig.3 and fig.4  we show the convergence of the excitation energy  for $Mg$ and $Si$
       isotopes respectively. For $Mg$ a  pattern  similar to $Ne$ can be seen.
\renewcommand{\baselinestretch}{1}
\begin{figure}
\centering
\includegraphics[width=10.0cm,height=10.0cm,angle=0]{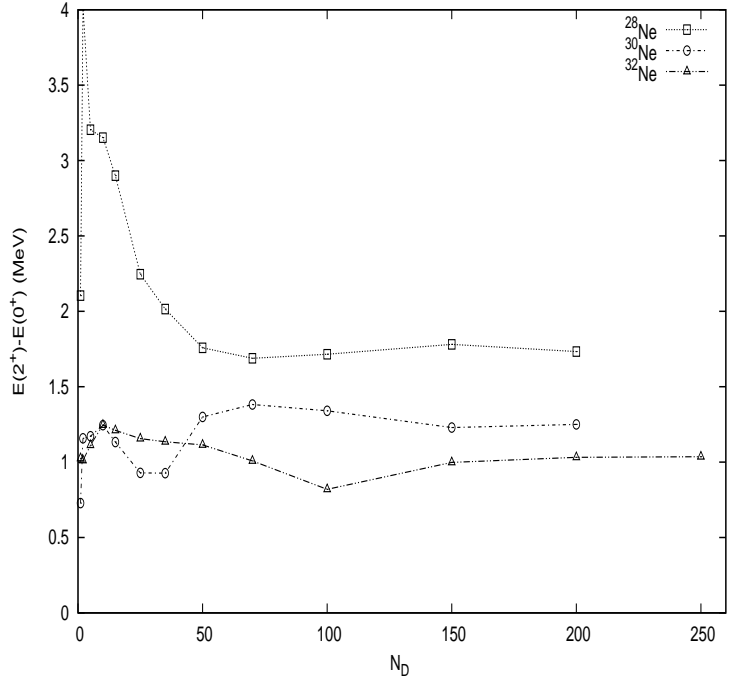}
\caption{Convergence of the excitation energy of $2^+_1$ as a function of the number of Slater
 determinants for $Ne$ isotopes.}
\end{figure}
\renewcommand{\baselinestretch}{2}
\renewcommand{\baselinestretch}{1}
\begin{figure}
\centering
\includegraphics[width=10.0cm,height=10.0cm,angle=0]{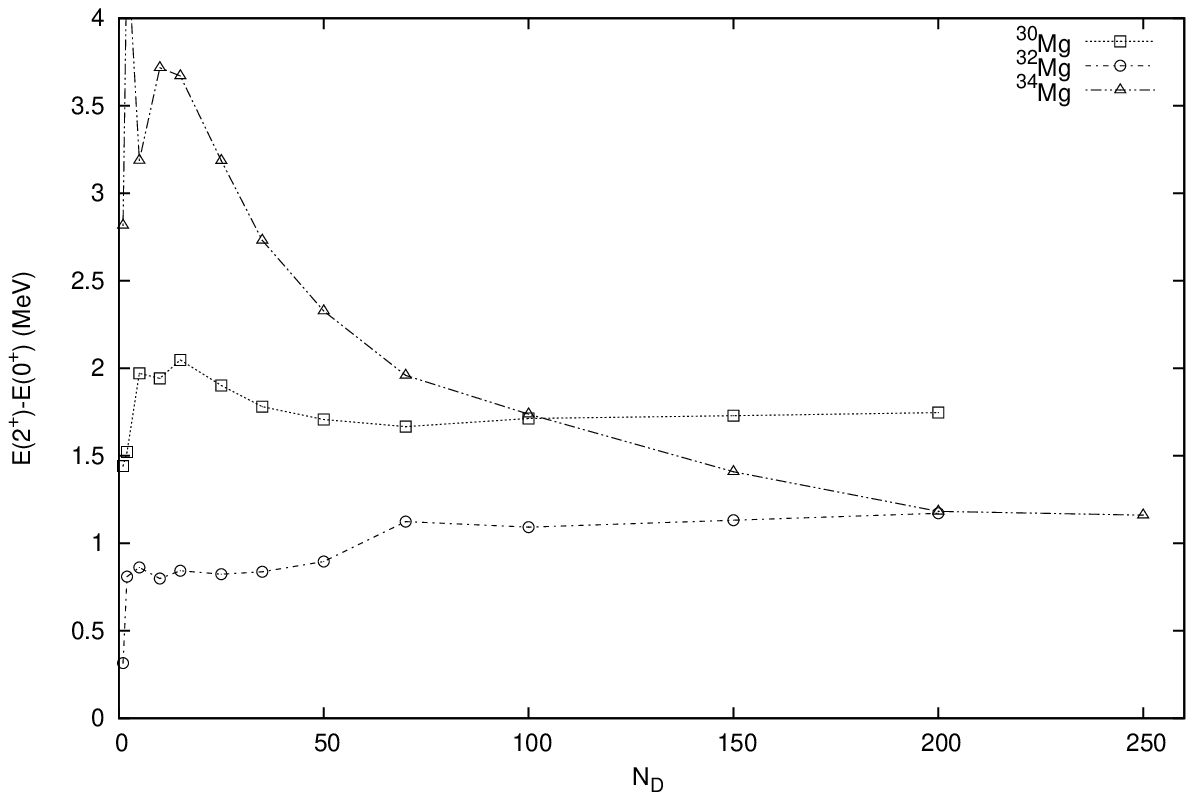}
\caption{Convergence of the excitation energy of $2^+_1$ as a function of the number of Slater
 determinants for $Mg$ isotopes.}
\end{figure}
\renewcommand{\baselinestretch}{2}
\renewcommand{\baselinestretch}{1}
\begin{figure}
\centering
\includegraphics[width=10.0cm,height=10.0cm,angle=0]{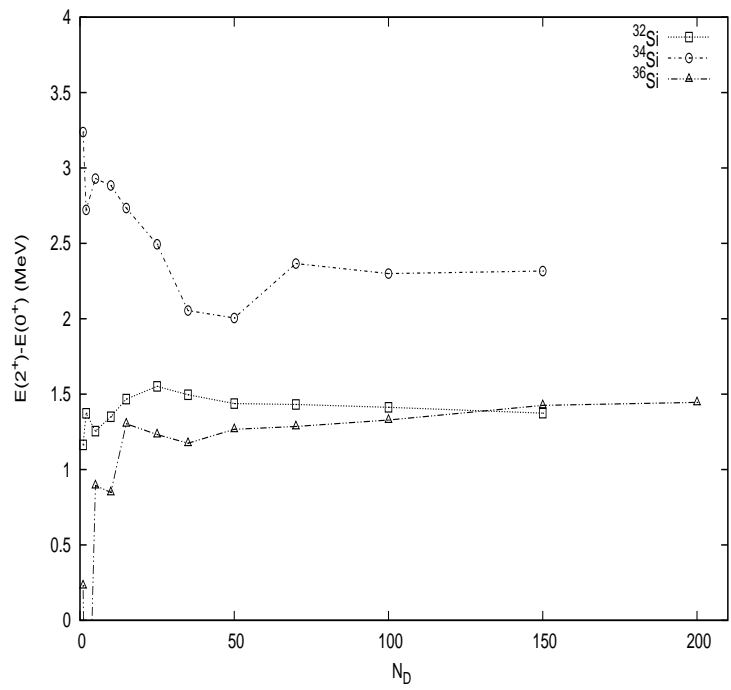}
\caption{Convergence of the excitation energy of $2^+_1$ as a function of the number of Slater
 determinants for $Si$ isotopes.}
\end{figure}
\renewcommand{\baselinestretch}{2}

       Notice the
       different behavior of $Si$ isotopes, where one can see an increase in the energy of the $2^+$ state
       for  $N=20$, again not to the same
       extent seen experimentally. We summarize the  results of the calculations in table 1 together
       with the corresponding experimental values of refs.[16],[17].
       It seems that the calculated values do not show
       enough collectivity. This is confirmed also by the calculated $BE2$ values which
       are systematically lower than the experimental values. 
       There could be several reasons for this, not necessarily related to the V18 interaction.
       First we renormalized to $5$ major shells only. Moreover the harmonic oscillator basis
       has the wrong large distance behavior instead of the proper exponential falloff. 
       For medium nuclei the h.o. representation might not be appropriate.
\renewcommand{\baselinestretch}{1}
\begin{table}
   \begin{tabular}{| c | c | c | c | c | c | }
          \hline
   Nucleus   & $  E(2^+)_{exp} $ & $  E(2^+)_{th}   $ & $ N_D   $ & $ BE2(th) $ & $ BE2(exp) $\\  
\hline
 $ {}^{28}Ne $ & $   1.304 $ & $  1.733$ & $ 200 $ & $   62.1   $ & $   136(23) $ \\
 $ {}^{30}Ne $ & $   0.792 $ & $  1.250$ & $ 200 $ & $   48.0   $ & $   226(35) $ \\
 $ {}^{32}Ne $ & $   0.722 $ & $  1.036$ & $ 250 $ & $   74.2   $ & $    -      $ \\
\hline
 $ {}^{30}Mg $ & $   1.4828$ & $  1.747$ & $ 200 $ & $   16.6   $ & $   273(26) $ \\
 $ {}^{32}Mg $ & $   0.8853$ & $  1.171$ & $ 200 $ & $   57     $ & $   434(52) $ \\
 $ {}^{34}Mg $ & $   0.660 $ & $  1.160$ & $ 250 $ & $   93.2   $ & $   573(79) $ \\
\hline
 $ {}^{32}Si $ & $   1.941 $ & $  1.373$ & $ 150 $ & $   120.8  $ & $122(+36 -21)$ \\
 $ {}^{34}Si $ & $   3.327 $ & $  2.316$ & $ 150 $ & $   37.9   $ & $   85(33) $   \\
 $ {}^{36}Si $ & $   1.408 $ & $  1.445$ & $ 200 $ & $   20.5   $ & $   193(59)$   \\
\hline
        \end{tabular}
 \caption { Experimental and calculated $E(2^+)$ in MeV and $BE2(0^+\rightarrow 2^+)$ in $e^2 fm^4$
            for Ne, Mg and Si isotopes. Experimental values are taken from refs. [16][17].}
\end{table}
\renewcommand{\baselinestretch}{2}
\renewcommand{\baselinestretch}{1}
\begin{table}
   \begin{tabular}{| c  c  c | c | c | c  |}
 \hline
 $ n $ & $ l $ & $  j $ & $    {}^{28}Ne$ & $ {}^{30}Ne$ & $ {}^{32}Ne$  \\
 \hline 
 $ 0 $ & $ 0 $ & $ 1/2 $ & $ 1.96 $ & $ 1.99 $ & $ 1.99 $  \\
 \hline
 $ 0 $ & $ 1 $ & $ 3/2 $ & $ 3.31 $ & $ 3.32 $ & $ 3.47 $  \\ 
 $ 0 $ & $ 1 $ & $ 1/2 $ & $ 1.82 $ & $ 1.84 $ & $ 1.85 $  \\ 
\hline
 $ 0 $ & $ 2 $ & $ 5/2 $ & $ 5.00 $ & $ 4.99 $ & $ 5.03 $  \\ 
 $ 0 $ & $ 2 $ & $ 3/2 $ & $ 2.06 $ & $ 3.74 $ & $ 3.76 $  \\ 
 $ 1 $ & $ 0 $ & $ 1/2 $ & $ 1.64 $ & $ 1.72 $ & $ 1.73 $  \\ 
\hline
 $ 0 $ & $ 3 $ & $ 7/2 $ & $ 0.08 $ & $ 0.08 $ & $ 0.50 $  \\ 
 $ 0 $ & $ 3 $ & $ 5/2 $ & $ 0.06 $ & $ 0.07 $ & $ 0.13 $  \\ 
 $ 1 $ & $ 1 $ & $ 3/2 $ & $ 0.65 $ & $ 0.66 $ & $ 1.77 $  \\ 
 $ 1 $ & $ 1 $ & $ 1/2 $ & $ 0.14 $ & $ 0.14 $ & $ 0.38 $  \\ 
\hline
 $ 0 $ & $ 4 $ & $ 9/2 $ & $ 0.04 $ & $ 0.03 $ & $ 0.05 $  \\ 
 $ 0 $ & $ 4 $ & $ 7/2 $ & $ 0.03 $ & $ 0.03 $ & $ 0.05 $  \\ 
 $ 1 $ & $ 2 $ & $ 5/2 $ & $ 0.83 $ & $ 0.93 $ & $ 0.89 $  \\ 
 $ 1 $ & $ 2 $ & $ 3/2 $ & $ 0.13 $ & $ 0.20 $ & $ 0.17 $  \\ 
 $ 2 $ & $ 0 $ & $ 1/2 $ & $ 0.26 $ & $ 0.26 $ & $ 0.25 $  \\ 
\hline
        \end{tabular}
 \caption { Neutron occupation numbers for $Ne$ isotopes.}
\end{table}
\renewcommand{\baselinestretch}{2}
\renewcommand{\baselinestretch}{1}
\begin{table}
        \begin{tabular}{| c  c  c | c | c | c  |}
                \hline
 $ n $ & $ l $ & $  j  $ & $    {}^{30}Mg$ & $ {}^{32}Mg$ & $ {}^{34}Mg$  \\ 
\hline 
 $ 0 $ & $ 0 $ & $ 1/2 $ & $   1.93 $ & $  1.99 $ & $  1.99 $ \\ 
\hline 
 $ 0 $ & $ 1 $ & $ 3/2 $ & $   3.29 $ & $  3.31 $ & $  3.46 $ \\
 $ 0 $ & $ 1 $ & $ 1/2 $ & $   1.80 $ & $  1.82 $ & $  1.83 $ \\
\hline 
 $ 0 $ & $ 2 $ & $ 5/2 $ & $   4.86 $ & $  4.90 $ & $  4.93 $ \\
 $ 0 $ & $ 2 $ & $ 3/2 $ & $   2.20 $ & $  3.64 $ & $  3.66 $ \\
 $ 1 $ & $ 0 $ & $ 1/2 $ & $   1.48 $ & $  1.67 $ & $  1.68 $ \\
\hline 
 $ 0 $ & $ 3 $ & $ 7/2 $ & $   0.08 $ & $  0.08 $ & $  0.57 $ \\
 $ 0 $ & $ 3 $ & $ 5/2 $ & $   0.06 $ & $  0.07 $ & $  0.12 $ \\
 $ 1 $ & $ 1 $ & $ 3/2 $ & $   0.66 $ & $  0.68 $ & $  1.76 $ \\
 $ 1 $ & $ 1 $ & $ 1/2 $ & $   0.16 $ & $  0.16 $ & $  0.35 $ \\
\hline 
 $ 0 $ & $ 4 $ & $ 9/2 $ & $   0.04 $ & $  0.03 $ & $  0.05 $ \\
 $ 0 $ & $ 4 $ & $ 7/2 $ & $   0.03 $ & $  0.03 $ & $  0.05 $ \\
 $ 1 $ & $ 2 $ & $ 5/2 $ & $   0.92 $ & $  1.02 $ & $  0.99 $ \\
 $ 1 $ & $ 2 $ & $ 3/2 $ & $   0.19 $ & $  0.30 $ & $  0.26 $ \\
 $ 2 $ & $ 0 $ & $ 1/2 $ & $   0.29 $ & $  0.32 $ & $  0.30 $ \\
\hline 
        \end{tabular}
 \caption { Neutron occupation numbers for $Mg$ isotopes.}
\end{table}
\renewcommand{\baselinestretch}{2}
\renewcommand{\baselinestretch}{1}
\begin{table}
        \begin{tabular}{| c  c  c | c | c | c  |}
                \hline
 $ n $ & $ l $ & $  j $ & $    {}^{32}Si$ & $ {}^{34}Si$ & $ {}^{36}Si $  \\
 \hline 
 $ 0 $ & $ 0 $ & $ 1/2 $ & $    1.86 $ & $ 1.99 $ & $ 1.99 $  \\
 $ 0 $ & $ 1 $ & $ 3/2 $ & $    3.29 $ & $ 3.29 $ & $ 3.49 $  \\
 $ 0 $ & $ 1 $ & $ 1/2 $ & $    1.80 $ & $ 1.80 $ & $ 1.81 $  \\
 \hline 
 $ 0 $ & $ 2 $ & $ 5/2 $ & $    4.76 $ & $ 4.82 $ & $ 4.85 $  \\
 $ 0 $ & $ 2 $ & $ 3/2 $ & $    2.50 $ & $ 3.54 $ & $ 3.57 $  \\
 $ 1 $ & $ 0 $ & $ 1/2 $ & $    1.21 $ & $ 1.62 $ & $ 1.63 $  \\
 \hline 
 $ 0 $ & $ 3 $ & $ 7/2 $ & $    0.07 $ & $ 0.06 $ & $ 0.53 $  \\
 $ 0 $ & $ 3 $ & $ 5/2 $ & $    0.06 $ & $ 0.06 $ & $ 0.11 $  \\
 $ 1 $ & $ 1 $ & $ 3/2 $ & $    0.68 $ & $ 0.70 $ & $ 1.85 $  \\
 $ 1 $ & $ 1 $ & $ 1/2 $ & $    0.17 $ & $ 0.19 $ & $ 0.30 $  \\
 \hline 
 $ 0 $ & $ 4 $ & $ 9/2 $ & $    0.04 $ & $ 0.02 $ & $ 0.03 $  \\
 $ 0 $ & $ 4 $ & $ 7/2 $ & $    0.03 $ & $ 0.03 $ & $ 0.04 $  \\
 $ 1 $ & $ 2 $ & $ 5/2 $ & $    0.99 $ & $ 1.12 $ & $ 1.09 $  \\
 $ 1 $ & $ 2 $ & $ 3/2 $ & $    0.25 $ & $ 0.40 $ & $ 0.36 $  \\
 $ 2 $ & $ 0 $ & $ 1/2 $ & $    0.30 $ & $ 0.37 $ & $ 0.35 $  \\
 \hline 
        \end{tabular}
 \caption { Neutron occupation numbers for $Si$ isotopes.}
\end{table}
\renewcommand{\baselinestretch}{2}
\renewcommand{\baselinestretch}{1}
\begin{table}
       \begin{tabular}{| c | c | c | c | c | c | c | c | c | c ||}
                \hline
$ N_{ho}$ & $ {}^{28}Ne$ & ${}^{30}Ne $ & $ {}^{32}Ne $ & $ {}^{30}Mg$ & ${}^{32}Mg $ & $ {}^{34}Mg $ & $
 {}^{32}Si$ & ${}^{34}Si $ & $ {}^{36}Si $ \\
\hline
$ 0 $ & $  1.96 $ & $  1.99 $ & $  1.99 $ & $ 1.93 $ & $  1.99 $ & $ 1.99$ & $  1.86 $ & $ 1.99 $ & $ 1.99 $ \\
$ 1 $ & $  5.13 $ & $  5.16 $ & $  5.32 $ & $ 5.09 $ & $  5.13 $ & $ 5.29$ & $  5.09 $ & $ 5.09 $ & $ 5.20 $ \\
$ 2 $ & $  8.7  $ & $ 10.45 $ & $ 10.51 $ & $ 8.54 $ & $ 10.21 $ & $10.27$ & $  8.47 $ & $ 9.98 $ & $ 9.95 $ \\
$ 3 $ & $  0.93 $ & $  0.95 $ & $  2.78 $ & $ 0.96 $ & $  0.99 $ & $ 2.8 $ & $  0.98 $ & $ 1.01 $ & $ 2.79 $ \\
$ 4 $ & $  1.29 $ & $  1.45 $ & $  1.41 $ & $ 1.47 $ & $  1.7  $ & $ 1.65$ & $  1.61 $ & $ 2.39 $ & $ 1.87 $ \\
\hline
       \end{tabular}
 \caption { Number of neutrons in the major h.o. shells for all cases.}
\end{table}
\renewcommand{\baselinestretch}{2}
\par
     It is instructive to evaluate the occupation numbers of neutrons that reveal which
     single-particle states are occupied. In tables 2-4. we show these occupation numbers for $Ne$,
     $Mg$ and $Si$ isotopes respectively.
     Notice that the $N_{ho}=4$ major shell is populated for all isotopes under
     consideration. The single-particle state $1d5/2$ has one neutron in all nuclei under consideration.
     Also the state $0p3/2$ has one neutron less than expected (the $N_{ho}=1$ core shell is excited).
     The $0f7/2$ orbital is nearly empty for neutron number $N=18,20$, and it is occupied
     (about $0.5$) only for $N=22$. This is the opposite of what has been found using realistic 
     effective interactions assuming an inert ${}^{16}O$ core and valence nucleons in the $sd-pf$ shells
     (ref. [5]). This is not necessarily in contradiction with the findings of ref.[3], since 
     our basis is expanded in an  spherical harmonic oscillator basis, while in ref. [3] 
     a deformed harmonic oscillator basis has been used. 
     A less detailed information can be obtained by evaluating
     the total number of neutrons in each major h.o. shell $N_{ho}$. The result is given in table 5.
     In all $Ne$ isotopes, the $N_{ho}=4$ major shell has about one neutron.
     Both ${}^{28}Ne$ and ${}^{30}Ne $ have one neutron in the  $N_{ho}=3$  fp shell and the added two neutrons
     occupy the $N_{ho}=2$ major shell. Adding two more neutrons, they mostly occupy the $N_{ho}=3$ shell.
     A similar pattern is seen also in $Mg$ isotopes. Silicon isotopes show a different
     behavior. Going from ${}^{32}Si$ to ${}^{36}Si$ one neutron occupies the 	$N_{ho}=3$ shell and 
     the other one occupies the $N_{ho}=4$ shell and two more neutrons occupy the $N_{ho}=3$ shell.
     Consider now the $N=20$ isotone chain. As we increase the number of protons the number of neutrons
     in the $N_{ho}=3$ shell does not change, instead the $N_{ho}=4$ becomes more populated at the expenses
     of the $N_{ho}=2$ shell.
     Let us remark that in this approach there are no single-particle energy levels. Moreover
     from fig.1, one can see that the amount of correlation energy is rather large, being about
     $15$ MeV's for ${}^{32}Mg$. 
\renewcommand{\baselinestretch}{1}
\begin{figure}
\centering
\includegraphics[width=10.0cm,height=10.0cm,angle=0]{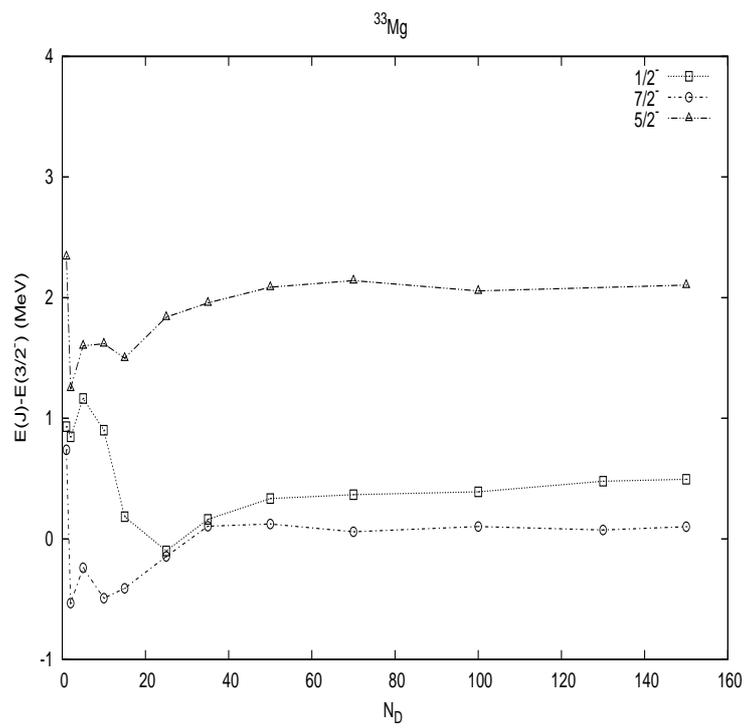}
\caption{Low energy levels of ${}^{33}Mg$ as a function of the number of Slater determinants.}
\end{figure}
\renewcommand{\baselinestretch}{2}
\par
     We have not performed a systematic study of even-odd nuclei. We studied only the case of ${}^{33}Mg$.
     For this nucleus we added a term $\gamma \HJ^2$ to the Hamiltonian with $\gamma=1 $MeV. Our calculation
     suggests that the spin of the ground state is $ 3/2^-$. The first excited state has $J^{\pi}=7/2^-$
     at an excitation energy of $0.1$MeV followed by a $1/2^-$ state at $0.49$MeV. The convergence curve
     as a function of the number of Slater determinants is shown in fig. 5. The spin of the ground state of this
     nucleus has been subject to some  debate (cf. ref.[25] and references in there). Our calculation shows that
     the main differences in the neutron occupation numbers are in the $fp$ shell. More explicitely,
     differences are seen for the  $3/2$ orbit   $n_{\nu}=1.4$, for the $1/2$ orbit $n_{\nu}=0.22$ and 
     for the $7/2$ orbit $n_{\nu}=0.13$. These values should be compared with the results of table 3. 
\bigskip
\section{ Concluding remarks.}
\bigskip
     A fully microscopic study of the islands of inversions, starting from
  a realistic bare NN interaction (such as the Argonne V18 interaction used in this work)
  is very challenging from a theoretical point of view.
 From one hand, the interaction needs to be renormalized to the adopted shell model space. 
 On the other hand, the shell model space cannot be too small if we are willing to consider
 only induced two-body interactions. Moreover the many-body
 technique used in solving the Schroedinger equation has to be capable of describing collective
 phenomena. Given the sizes of the Hilbert space, we have selected in this work to use the
 HMD method, whereby the nuclear wave function is approximated as a linear combination
 of a large number of angular momentum and parity projected Slater determinants. Although
 the results show a melting of the $N=20$ shell closure for $Ne$ and $Mg$ isotopes,
 and a restoration of shell closure for $Si$ isotopes, some issues remain open, most notably 
 we obtain too small $E2$ transition rates. With respect to this problem, the recently proposed 
 single-particle basis of ref. [21], might be useful in clarifying this issue.
 In the future we plan to use different NN interactions as the CDBonn2000 (ref. [30]) or the chiral
 N3LO (ref.[31]) and different renormalization prescriptions to investigate their predictions
 regarding this island of inversion.
\vfill
\section{ Appendix}
\bigskip
{\it {The Lee-Suzuki renormalization}}.
\bigskip
\par
       The nuclear Hamiltonian is
$$
H= \sum_{i=1}^A {p^2_i\over 2m}+\sum_{i<j} V^{NN}_{ij}
\eqno(A1)
$$
       where $m$ is the nucleon mass which we take equal to twice the reduced mass
       and $ V^{NN}_{ij}$ is the Argonne $v18$ interaction.
       To this Hamiltonian, similarly to what is done in the no core shell model
       approach (NCSM) (ref. [11]), we add a center of mass harmonic potential
$$
V_{cm}= {1\over 2} m A \om^2 R^2_{cm}
\eqno(A2)
$$
       The resulting Hamiltonian $H^{\om,A}$ can be rewritten as
$$
H^{\om,A}=\sum_{i=1}^A h_i +\sum_{i<j} V^{\om,A}_{ij}
\eqno(A3)
$$
       where 
$$
h_i = {p^2_i\over 2 m} + {1\over 2} m \om^2 r^2_i 
\eqno(A4a)
$$
      and
$$
 V^{\om,A}_{ij}= V^{NN}_{ij}- {m \om^2 \over 2 A} (\vec r_i-\vec r_j)^2
\eqno(A4b)
$$
       At the cluster-2 approximation we consider the 2-particle Hamiltonian
$$
H^{\om,A}_2 = h_1+h_2 +V^{\om,A}_{12}
\eqno(A5)
$$
       This Hamiltonian separates into a center of mass Hamiltonian for the 2 particles
       plus a  Hamiltonian in the relative coordinates $r, p$
$$
H_r  = {p^2\over m} + {1\over 4} m \om^2 r^2+ V^{\om A}_{12}\equiv H_{0r}+V^{\om A}_{12}
\eqno(A6)
$$
       which defines $H_{0r}$.
       The Hamiltonian $H_r$ of relative motion is the one that is renormalized using 
       the Lee-Suzuki procedure.
       For all values of the spin, the angular momentum, the isospin and the z-projection
       of the isospin, $s,j,t$ and $t_z$ all matrix elements of this bare
       Hamiltonian are evaluated (the radial quantum number can reach $200\div 250$).
       The radial quantum number must be large enough so that the single-particle
       space is complete for all practical purposes.
       Let us call $H_{ij}$ the resulting matrix. The indices $i,j$ are defined in
       the full single-particle space of relative coordinates.
       The renormalization procedure consists
       in separating the full space  in two parts with the aid
       of projectors $P$ (model space) and $Q$ (excluded space, such that 
       $P+Q=1$. According to the UMOA prescription a unitary transformation
       is performed on $H$ such that there is no coupling between the P space and the Q
       space. Briefly summarized the method consists in the following steps. 
       Let $p,p'$ be indices in the model space
       and $q,q'$ be the indices of the excluded space. First we diagonalize the bare
       Hamiltonian and let $ H= V \eps \tilde V$ the corresponding eigenvalue
       equation ($\eps$ and $V$ are the energy eigenvalues and eigenvectors). 
       Let us call $ U_{p,p'}= V_{p,p'}$ and $W_{q,p}=V_{q,p}$ 
       the P-part and the QP-part of the matrix of the eigenvectors $V$, respectively, and
       let
$$
S_{q,p} =\sum_{p'} W_{q,p'}U^{-1}_{p'p}
\eqno(A7)
$$
       where the sum runs over the P-indices. We avoid the use of the traditional symbol
       $\om$ for this matrix, in order not to confuse it with the h.o. frequency.
       Let us also construct the P-space  matrix
$$
N_{p,p'} =  \delta_{p,p'}+\sum_q  S_{q, p}\; S_{q, p'}
\eqno(A8)
$$
       $\delta$ being the Kronecker $\de$.
       Further, let us build the matrix
$$
\Om_{p,p'}=N^{-1/2}_{pp'},\;\;\;\;\;\Om_{q,p}=\sum_{p'}S_{q,p'}N^{-1/2}_{p'p}
\eqno(A9)
$$
       which connects the $P+Q$ space with the $P$ space.
       Then the renormalized Hamiltonian is then given by
$$
H^{ren}_{pp'} = \sum_{ij}\Om_{i,p}H_{ij}\Om_{j,p'}
\eqno(A10)
$$
       where in the above equation the sum is over the full space.
       The eigenvalues of $H^{ren}$ coincide (almost to machine accuracy)
       with the P-part of the eigenvalues of the bare $H$. The full details
       of the proof can be found in ref. [7]-[10]. 
\par
       Note however that recently the above prescription has been recast
       in a more simplified and transparent form in ref. [26]. The two formulations
       can be shown identical using the property $VPV^{\dagger}P= PVPV^{\dagger}P+
       QVPV^{\dagger}P= UU^{\dagger}+ W U^{\dagger}$.
       This renormalization prescription can be formulated in terms of the singular
       value decomposition of the matrix $U$ (ref. [27]), which is numerically 
       very robust. The main point of ref.[26] and ref. [27] is to rewrite the renormalized Hamiltonian
       as
$$
H^{ren}= (UU^{\dagger})^{-1/2} U (P\eps) U^{\dagger}(UU^{\dagger})^{-1/2}	
\eqno(A11)
$$
       and (cf. ref. [27])
$$
(UU^{\dagger})^{-1/2}	U= X Y^{\dagger}
\eqno(A12)
$$
       where $X $ and $Y^{\dagger}$ are the left and right singular eigenvectors of $U$.
\par\noindent
       The P-indices are relative to major harmonic oscillator shells.
       Typically we renormalize to $N_{ren}=8$  major shells in the intrinsic
       frame. We point out that these are not the major shells used in the
       variational calculation as discussed later.
\par
       Once we obtain the renormalized 2-particle Hamiltonian we can define an effective
       potential for two particles from $H^{ren}$ as
$$
V^{ren} = H^{ren}-H_{0r}
\eqno(A13)
$$
       This interaction replaces $V^{\om,A}_{ij}$ in eq.(A3). Next we subtract from 
       eq(A3) the Hamiltonian of the center of mass in order to obtain the intrinsic Hamiltonian
       and the final result is, for the 2 particle system,
$$
H^{int}_{1,2}= H_{0r}+V^{ren} + ( {2\over A}-1)H_{0r}
\eqno(A14a)
$$
       and
$$
H^{int}=\sum_{i<j} H^{int}_{i,j}
\eqno(A14b)       
$$
       for the A-particle system.
       We are now in a position to transform these matrix elements from the
       intrinsic frame to the lab frame, using the Talmi-Moshinsky (cf. ref. [28] and references in there
 for a very efficient implementation)
      transformation brackets,
\par\noindent
       To the final Hamiltonian we add the term
$$
\beta ( H_{cm} -3 \hbar \om /2)
\eqno(A15)
$$
       with $\beta>0$ to prevent center of mass excitations.
       We end up with all possible matrix elements 
       $<n_a l_a j_a n_b l_b j_b J|H|n_c l_c j_c n_d l_d j_d J>$ in the lab frame for
       the $nn,np$ and $pp$ interaction. The quantum numbers in the lab frame,
       satisfy  
$$
2 n_a+ l_a +2 n_b+ l_b \leq N_{ren}
\eqno(A16)
$$.
\par
       Most important, we consider
       quantum numbers $n,l$ such that
$$
2 n+l\leq N_{ren}/2
\eqno(A17)
$$
 That is, we use an  energy  truncation scheme, called HMD-a in refs. [14],[15]. The use of condition
  (A16) is called HMD-b in refs.[14],[15]
\par\noindent
       As discussed in ref. [29], there are many ways to renormalize
       the Hamiltonian. The one discussed above is the one adopted in all calculations
       of this work.
       Let us remark that instead of condition (A17) we could have adopted the 
       condition (A16) for the quantum numbers in the lab frame. We prefer condition (A17) since
       (A16) strongly overbounds unless we consider many major shells. Condition (A16) 
       is useful as a numerical test for the deuterium.
       We have performed a numerical test using $N_{ren}=6$ ($7$ major shells in the lab frame)
       with $\hbar\om=12$MeV for the deuterium. The HMD-b method for $5$ Slater determinants
       gave a discrepancy from the exact binding energy of the deuterium of $0.27$KeV. 
       Using $10$ Slater determinants this discrepancy has been reduced to $1.3$eV and using
       $15$ Slater determinants this discrepancy has been further reduced to $0.015$eV.
       By exact binding energy we mean the value obtained by diagonalizing the bare  Hamiltonian
       matrix in the intrinsic frame  with 480 major oscillator shells.

\vfill
\eject
\end{document}